\documentclass[prl,twocolumn,floatfix,superscriptaddress,nofootinbib,aps]{revtex4}

\usepackage{natbib}
\usepackage{bm}

\usepackage{graphicx}
\usepackage{amsmath}
\usepackage{amssymb}
\usepackage{hyperref}
\usepackage{verbatim}

\font \bolditalics = cmmib10
\def\bx#1{\leavevmode\thinspace\hbox{\vrule\vtop{\vbox{\hrule\kern1pt
        \hbox{\vphantom{\tt/}\thinspace{\bf#1}\thinspace}}
      \kern1pt\hrule}\vrule}\thinspace}
\def \vc #1{{\textfont1=\bolditalics \hbox{$\bf#1$}}}

\def\kg{{\bf k}}

\def\thetavc{{\vc \theta}}

\begin{document}

\title{Detection of warm and diffuse baryons in large scale structure from the cross-correlation of gravitational lensing and the thermal Sunyaev-Zeldovich effect}

\author{Ludovic Van Waerbeke}
\email{waerbeke@phas.ubc.ca}
\affiliation{University of British Columbia, Department of Physics and Astronomy, 6224 Agricultural Road, V6T 1Z1, Vancouver, Canada}

\author{Gary Hinshaw}
\email{hinshaw@phas.ubc.ca}
\affiliation{University of British Columbia, Department of Physics and Astronomy, 6224 Agricultural Road, V6T 1Z1, Vancouver, Canada}
\affiliation{Canada Research Chair in Observational Cosmology}

\author{Norman Murray}
\email{murray@cita.utotonto.ca}
\affiliation{Canadian Institute for Theoretical Astrophysics, University of Toronto, 60 St. George street, M5S 3H8, Toronto, Canada}
\affiliation{Canada Research Chair in Theoretical Astrophysics}


\begin{abstract}
We report the first detection of a correlation between gravitational lensing by large scale structure and the thermal Sunyaev-Zeldovich (tSZ) effect. Using the mass map from the Canada France Hawaii Telescope Lensing Survey (CFHTLenS) and a newly-constructed tSZ map from Planck, we measure a non-zero correlation between the two maps out to one degree angular separation on the sky, with an overall significance of $6\sigma$.  The tSZ maps are formed in a manner that removes primary cosmic microwave background fluctuations and minimizes residual contamination by galactic and extragalactic dust emission, and by CO line emission.  We perform numerous tests to show that our measurement is immune to these residual contaminants. The resulting correlation function is consistent with the existence of a warm baryonic gas tracing the large scale structure with a bias $b_{\rm gas}$.  Given the shape of the lensing kernel, our signal sensitivity peaks at a redshift $z\sim 0.4$, where half a degree separation on the sky corresponds to a physical scale of $\sim$10 Mpc. The amplitude of the signal constrains the product $(b_{\rm gas}/1)(T_e / 0.1 \textrm{ keV})(\bar{n}_e / 1\textrm{ m}^{-3})=2.01\pm 0.31\pm 0.21$, at redshift zero. 
Our study suggests that a substantial fraction of the ``missing'' baryons in the universe may reside in a low density warm plasma that traces dark matter.
\end{abstract}

\maketitle

\section{Introduction}
\label{s:intro}
A crucial element in the study of structure formation is to understand how the baryons interact with dark matter.
Approximately 10\% of the baryons are in compact objects (mostly stars and dust) while $\sim$90\% are in the form of a diffuse gas component. Due to this extreme diversity of physical states, the observation of baryons requires multi-wavelength measurements. Dark matter on the other hand can only be mapped with the gravitational lensing effect \cite{1984ApJ...281L..59T,1991ApJ...380....1M, Kaiser92}. Better observational constraints on the interaction of baryons and dark matter will advance our understanding on two distinct fronts: first, it will improve our understanding of baryonic processes in structure formation; second it will enhance the utility of gravitational lensing as a cosmological probe down to galactic scales, where baryonic processes pose a challenge to the interpretation of lensing data. 
Within the hierarchical structure formation scenario, galaxy-galaxy lensing is successfully probing the connection between dark matter haloes and the stellar mass distribution in various galaxy types \citep{2006MNRAS.368..715M, 2012ApJ...744..159L}, but these studies only probe the 10\% of baryons in compact objects.  Sensitivity to the dominant diffuse component requires probing larger scales: groups and clusters of galaxies. Unfortunately, this gas is difficult to observe, except in the densest regions where its temperature is high enough to be seen with X-ray or Sunyaev-Zeldovich measurements \citep{2011ARA&A..49..409A}.

A comparison of group and cluster masses derived from dynamical and X-ray data reveals that baryons are missing at all scales \citep{2010ApJ...719..119D}; the problem is particularly severe at galactic halo scales. This is likely connected to the missing baryon problem occurring at redshift $z<2$, where the inter-cluster gas becomes ionized in a warm phase that is particularly difficult to observe with current facilities.  Numerical simulations have predicted long ago that the missing baryons could be in a low-density warm plasma ($\sim$10$^5-10^7$ K) associated with large structures, or even filaments \citep{1996ApJ...464...27O,1999ApJ...514....1C,2001ApJ...552..473D,2008AN....329..118B}. There is indeed some evidence for an intergalactic warm plasma, but the data interpretation is highly model dependent \citep{1993ApJ...409L..37W,2009ApJ...695.1127G}. The fact that $\sim$50\% of the baryons are in a physical state largely unconstrained by observations, and that the processes responsible for this situation are not understood, poses a challenge to the theory of structure formation. Recently, it has also been realized that it poses a problem for the interpretation of gravitational lensing because baryonic processes could impact the dark matter distribution, even on large scales, via gravitational feedback \citep{Semboloni11}.




The thermal Sunyaev-Zeldovich (tSZ) effect \citep{1972CoASP...4..173S} offers a unique observable for probing the diffuse baryonic component; unlike X-ray luminosity, the tSZ signal is linearly proportional to the baryon density, making the detection of low-density gas more feasible.  Moreover, the tSZ signal strength is independent of redshift, making it an especially useful observable to cross-correlate with signals more localized in redshift. For instance, the diffuse baryonic component can be inferred  by the cross-correlation between tSZ and galaxy surveys. Recently the Planck Collaboration performed such analysis on the Sloan Digital Sky Survey maxBCG clusters catalogue \citep{2007ApJ...660..239K}; they reported a tSZ decrement \citep{2011A&A...536A..12P} significantly lower than expected, a finding confirmed by \citep{2012PhRvD..85b3005D}. This suggests that either the modeling of the intra-cluster medium and/or the selection of optical clusters is not understood well enough. 

These inconclusive results show that the interpretation of the cross-correlation between tSZ and large scale structures necessitates an unbiased tracer of large scale structure mass.  Gravitational lensing provides an unbiased tracer of the projected mass, independent of its dynamical and physical state.  Cross-correlating tSZ and gravitational lensing data was proposed in \citep{2000ApJ...540....1C,2000PhRvD..62j3506C}, but suitable data sets of each component have only recently become available.  In this paper we present a cross-correlation analysis between a new Planck tSZ map and the Canada France Hawaii Telescope Lensing Survey (CFHTLenS) mass map \citep{2013MNRAS.433.2545E} and investigate the implications for warm baryons.

\section{Formalism}
\label{s:mech}

The formalism for correlating the gravitational lensing and tSZ signals are introduced here.  For the lensing signal it is useful to define the kernel $W^\kappa(w)$, which contains the geometrical factor that all lensing quantities depend on. It is given by:
\begin{equation}
W^\kappa(w)={3\over 2}\Omega_0 \left({H_0\over c}\right)^2 g(w) {f_K(w)\over a},
\end{equation}
where $w(z)$ is the comoving radial distance at redshift $z$, $f_K(w)$ is the corresponding angular diameter
distance, $a=1/(1+z)$ the scale factor, and $g(w)$ is a function which depends on the redshift distribution of
the sources $p_S(w)$:
\begin{equation}
g(w)=\int_w^{w_H}{\rm d}w' p_S(w'){f_K(w'-w)\over f_K(w')},
\end{equation}
where $w_H$ is the distance to the horizon. The projected mass density along a line-of-sight specified by
position angle $\thetavc$ on the sky is characterized by the convergence $\kappa(\thetavc)$:
\begin{equation}
\kappa(\thetavc)=\int_0^{w_H} {\rm d} w W^\kappa(w) \delta_m(f_K(w)(\thetavc),w),
\label{kappamapdef}
\end{equation}
where $\delta_m(f_K(w)(\thetavc),w)$ is the 3-dimensional mass density contrast. The power spectrum of the
convergence map $C_l^\kappa$ is obtained using the Limber approximation \citep{Kaiser92}:
\begin{equation}
C_l^\kappa=\int_0^{w_H} {\rm d} w \left[{W^\kappa(w)\over f_K(w)}\right]^2 P_{mm}\left({l\over f_K(w)},w\right),
\label{kappaCel}
\end{equation}
where $P_{mm}(k)$ is the power spectrum of the 3-dimensional mass density $\delta_m$ defined in the usual way:
\begin{equation}
\langle \tilde\delta(\kg)\tilde\delta^\star(\kg')\rangle=(2\pi)^3 \delta_K(\kg-\kg') P_{mm}(k).
\end{equation}
%
%
%

The tSZ effect is produced by inverse Compton scattering of Cosmic Microwave Background (CMB) photons by
hot relativistic electrons. At frequency $\nu$, the tSZ-induced temperature change, $\Delta T$, along the line-of-sight
is characterized by the tSZ Compton parameter $y$:
\begin{equation}
{\Delta T\over T_0}=y \, S_{\rm SZ}(x),
\end{equation}
where $S_{\rm SZ}(x) = x\coth (x/2) -4$ is the tSZ spectral dependence, given in terms of $x=h\nu/k_B T_0$.  Here $h$ is the Planck constant, $k_B$ is the Boltzmann constant, and $T_0=2.725$ K is the CMB temperature \citep{2009ApJ...707..916F}. The Comptonization parameter $y$ is given by the line-of-sight integral of the electron pressure:
\begin{equation}
y(\thetavc) = \int_0^{w_H} a{\rm d}w ~ {k_B \sigma_T\over m_e c^2} n_e T_e,
\end{equation}
where $\sigma_T$ is the Thomson cross-section, and $n_e$ and $T_e$ are the electron number density and temperature respectively. 

In this paper we adopt a simple constant bias model where the gas density is proportional to the mass density with a proportionality factor that depends only on redshift. Following \cite{1999PhRvD..59j3002G} the gas mass density is given by $\delta_{\rm gas}=b_{\rm gas}(z)\delta_{\rm mm}$ with $b_{\rm gas}(z)\propto (1+z)^{-1}$ and $\delta_{mm}$ being the mass density contrast. In this model, gas temperature fluctuations are ignored and the average temperature is proportional to $(1+z)^{-1}$. It follows that the pressure and matter power spectra, $P_{\pi\pi}$ and $P_{\rm mm}$, and their cross-spectrum, $P_{\pi m}$, are simply related via a pressure bias $b_\pi$:
\begin{eqnarray}
P_{\pi\pi}(k)&=&b_\pi^2 P_{mm}(k) \nonumber \\
P_{\pi m}(k)&=&b_\pi P_{mm}(k).
\end{eqnarray}
Hydrodynamic numerical simulations \citep{1995ApJ...442....1P,2000ApJ...531...31R,2001PhRvD..63f3001S} suggest that a constant bias $b_\pi$ is an acceptable approximation down to $k\sim 3~h Mpc^{-1}$, approximately scaling as $(1+z)^{-1}$ \citep{1995ApJ...442....1P,1999PhRvD..59j3002G}. This model is sufficiently accurate for interpreting the initial detection presented below; a more sophisticated approach, based on the halo model, is deferred to future work. Since our study is only sensitive to scales larger than a typical cluster of galaxies, we will ignore $k$ dependence in the bias and treat $b_\pi$ as a free parameter proportional to the scale factor $a$ \citep{1999PhRvD..59j3002G}. Given these assumptions, the pressure bias can be written as:
\begin{equation}
b_\pi(z)={b_\pi(0)\over 1+z}={b_{\rm gas}(0)\over 1+z} \,  {k_B T_e(0)\over m_e c^2}
\end{equation}

The Sunyaev-Zeldovich kernel,  $W^{SZ}$, can be defined as
\begin{equation}
W^{\rm SZ}=\bar{n}_e \sigma_T \, {k_B T_e(0)\over m_ec^2} \, b_{\rm gas}(0),
\end{equation}
where $\bar{n}_e$ is the average electron number density today. It is convenient to rewrite this as
\begin{equation}
W^{\rm SZ}=0.00196~\sigma_T \, b_{\rm gas} \, \left({T_e(0)\over 1\textrm{ keV}}\right)\left({\bar{n}_e\over1\textrm{ m}^{-3}}\right).
\label{WSZdef}
\end{equation}
Using the above simple bias model, the tSZ-$\kappa$ angular cross-power spectrum can then be written as \cite{2000ApJ...540....1C}:
\begin{equation}
C_l^{{\rm SZ}-\kappa}=\int_0^{w_H} {\rm d} w \left[{W^{\rm SZ}(w)W^\kappa(w)\over f_K^2(w)}\right]
P_{mm}\left({l\over f_K(w)},w\right).
\label{SZkappaCel}
\end{equation}
The cross-correlation function in real space, $\xi_{y\kappa}(r)$, is given by the sum:
\begin{equation}
\xi_{y\kappa}(r)={1\over 4\pi}\sum_l (2l+1) C_l^{{\rm SZ}-\kappa}J_0(lr)W_l(\theta_0)W_l(\theta_{\rm SZ}),
\label{SZkappaXi}
\end{equation}
where $W_l(\theta_0)$ and $W_l(\theta_{\rm SZ})$ are the smoothing window functions of the lensing and tSZ maps, respectively.

%
%

\section{The $\kappa$ and $y$ maps}

\begin{figure}
\includegraphics[scale=0.5]{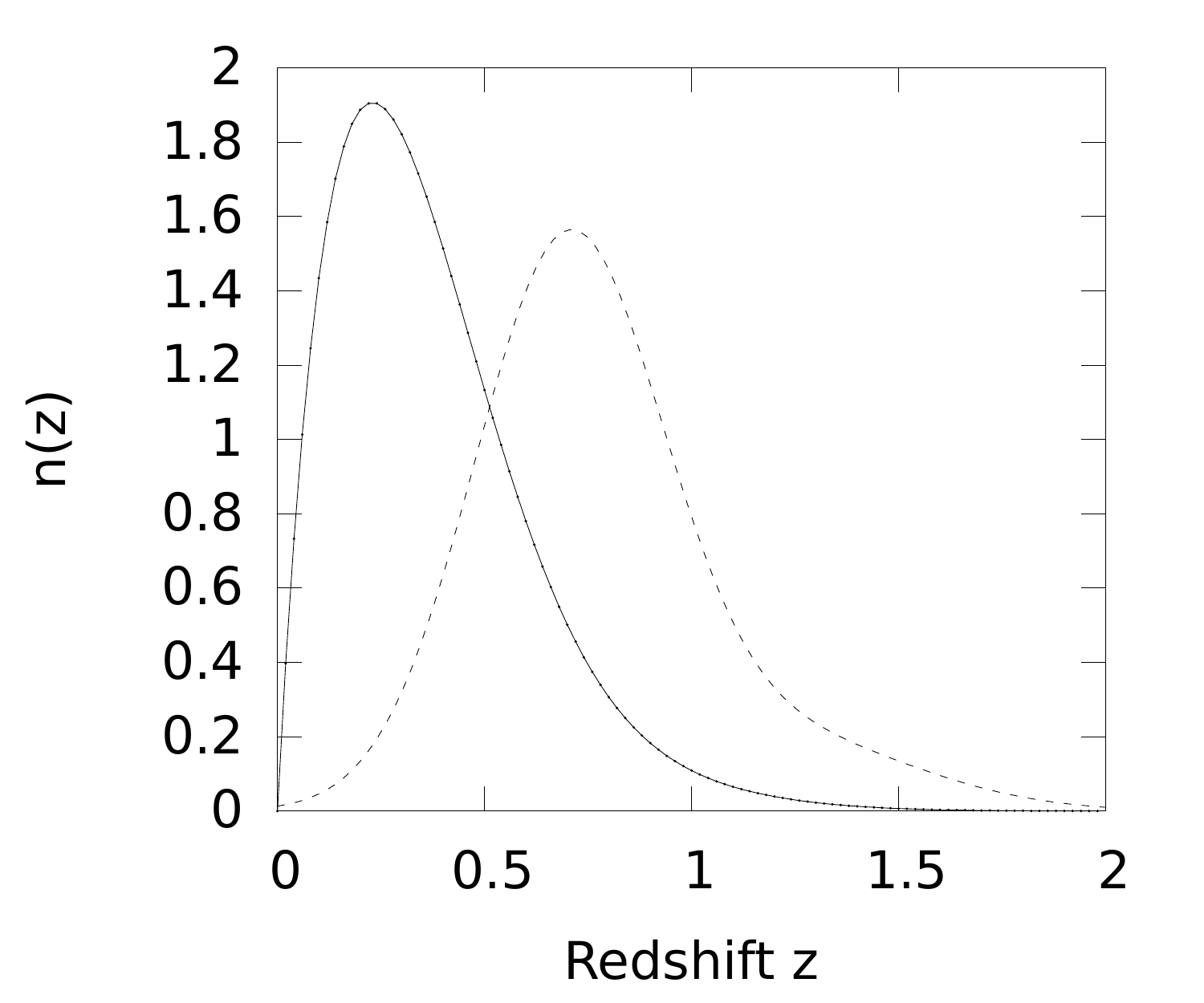}
\caption{\label{nofz} The dashed line shows the redshift distribution of the sources used in the mass reconstruction in the lensing map $\kappa(\thetavc)$. The solid line shows the efficiency of the lenses integrated over this source distribution; it peaks at a redshift of $z\sim 0.37$.  The $\kappa$ map is dominated by lenses located in the $z=[0,0.5]$ redshift range.}
\end{figure}

The gravitational lensing maps used in this study are from the Canada France Hawaii Telescope Lensing Survey (CFHTLenS) \citep{2013MNRAS.433.3373V}. They cover a total of $154$ deg$^2$ in four compact regions, W1, W2, W3 and W4, corresponding to an area on the sky of approximately $72$, $36$, $50$ and $25$ deg$^2$ respectively \citep{2013MNRAS.433.2545E}. The lensing signal measured in the maps is consistent with the predictions for large-scale structure in the LCDM cosmology, and they have been tested against all known sources of residual systematics. The mass maps can therefore be used without further testing and we direct the reader to \citep{2013MNRAS.433.3373V} for a thorough description of the mass reconstruction procedure, the testing, and the cosmological analysis of the maps. For the purpose of our study the lensing map $\kappa$ is smoothed with a Gaussian window of width $\theta_0=6~{\rm arcmin}$. As a sanity check, we also generate a lensing map in which all source galaxies have been rotated by $45$ degrees. The cosmological lensing signal in these maps is suppressed by the rotation; these maps are called the $B$-mode maps. Figure~\ref{nofz} shows the source redshift distribution (dashed line) used for the mass reconstruction.  The lensing signal, and the resulting mass reconstruction, is most sensitive in the redshift range approximately half way between the sources and the observer. The solid line in Figure~\ref{nofz} shows the lens redshift sensitivity characterized by the ratio $W^\kappa(w)/f_K(w)$ in Eq.~\ref{kappaCel}.  Since the tSZ signal strength is independent of redshift, cross-correlation of the lensing mass map with the tSZ map will roughly track this redshift distribution.

For the tSZ signal, multiple full-sky maps of the Comptonization parameter, $y$, were constructed from the Planck data by forming linear combinations of the four lowest-frequency band HFI maps: 100 GHz, 143 GHz, 217 GHz, and 353 GHz.  The band coefficients are chosen to project out contaminating signals, as described below.  The band maps used at each frequency were the nominal 15-month combined-survey maps in thermodynamic CMB temperature units, smoothed to a gaussian FWHM beam resolution of 9.5 arcmin.  The dominant emission components in this frequency range are 
\begin{enumerate}
\item Primary CMB fluctuations, with a constant frequency spectrum in thermodynamic units.
\item Thermal Sunyaev-Zeldovich fluctuations, with frequency dependence $\Delta T_{\rm SZ}(\nu)/T_0 = y \, S_{\rm SZ}(x)$, where $T_0 = 2.725$ K is the monopole temperature, $y$ is the Comptonization parameter along the line-of-sight, and $S_{\rm SZ}(x) = x \coth (x/2) - 4$, with $x \equiv h\nu/k_BT_0$.
\item Radio emission from the interstellar medium; the dominant spectral component above 100 GHz is free-free emission from classical H II regions and the warm ionized medium, with a frequency spectrum $T^A_{\rm ff}(\nu) \propto \nu^{-2.15}$ in antenna temperature units.
\item Line emission from interstellar CO molecules, which emit in the 100, 217, and 353 GHz bands; we discuss line strength ratios below.
\item Thermal emission from interstellar dust grains and from dusty external galaxies (including the Cosmic Infrared Background, CIB), with a frequency spectrum $T^A_{\rm d}(\nu) \propto \nu^{\beta_d}$, where $\beta_{\rm d} \sim 1.8$ in antenna temperature units.
\end{enumerate}

\begin{table}
\caption{Band data for the Planck $y$ maps\label{tab:sz_coeff}}
\begin{tabular}{lrrrrrrr}
\hline
Map & $\beta_{\rm d}$ & $b_{100}$ & $b_{143}$ & $b_{217}$ & $b_{353}$  & $r_{2.0}$ \\
\hline
A      & $\cdots$& $-$0.1707 & $-$0.1148 & 0.0085 & 0.2770  & 44.42 \\
B      & 1.4 & $-$0.7089 & $-$0.1372 & 0.9388 & $-$0.0927  & $-$1.99 \\
B$'$\tablenotemark[1]  & 1.4 & $-$1.2749 & 0.7160  &  0.6374  & $-$0.0768  & $-$2.94 \\
C      & 1.6 & $-$0.6952 & $-$0.1378 & 0.9169 & $-$0.0839  & $-$3.13 \\
D      & 1.8 & $-$0.6826 & $-$0.1385 & 0.8969 & $-$0.0758  & $-$0.95 \\
D$'$\tablenotemark[1] & 1.8 & $-$1.2236 & 0.6750  &  0.6118  & $-$0.0632  & $-$0.89 \\
E      & 2.0 & $-$0.6710 & $-$0.1393 & 0.8787 & $-$0.0684  & 0.00 \\
E$'$\tablenotemark[2] & 2.0 & 0.0113 & 0.0006 & $-$0.0196 & 0.0078  & 1.00  \\
G     & 1.0 & $-$0.7389 & $-$0.1364 & 0.9876 & $-$0.1124  & $-$5.67 \\
H\tablenotemark[3]     & 1.6 & 0.6483 & $-$2.1607 & 1.6287 & $-$0.1163 & $-$2.31 \\
I       & $\cdots$ & $\cdots$ & $-$0.9656 & 0.9656 & $\cdots$ & 10.38 \\
\hline
$c_{\nu}$\tablenotemark[4] & & 1.288 & 1.657 & 3.003 & 13.012  & \\
$S_{\rm SZ}(x)$ & & $-$1.506 & $-$1.037 & $-$0.001 & 2.253  & \\
\hline
\end{tabular}
\label{tbl:var}
\tablenotetext[1]{Maps B$'$ and D$'$ apply the same spectral conditions as maps B and D, respectively, but they additionally project out CO line emission as described in the text.}
\tablenotetext[2]{Map E$'$ is the complement of map E: it projects {\em out} tSZ emission while retaining a dust component with spectral index $\beta_{\rm d}=2.0$, normalized at 100 GHz.}
\tablenotetext[3]{Map H applies the same conditions as map C, but additionally projects out radio emission with spectrum $\beta_{\rm ff}=-2.15$.}
\tablenotetext[4]{The conversion factor from antenna to thermodynamic temperature.}
\end{table}

The spectra of the CO and dust emission vary from pixel to pixel in the map, and span a range of spectra in the vicinity of the nominal spectra described above, so the $y$ maps we construct will inevitably have residual contamination from unprojected signal components.  However, by construction, the level of residual contamination varies widely among the different estimates we form.  Further, given the band coefficients used in any given estimate, we can {\em predict} the fractional foreground residual as a function of the ``true'' spectral index, as discussed below.   We can thus place bounds on residual foreground contamination by examining the dependence of the SZ-lensing cross-correlation on the set of $y$ estimates we generate.  In practice, we find a very weak dependence, which will allow us to place useful bounds on the level of contamination that correlates with the lensing mass map.

\begin{figure*}
\begin{center}
\includegraphics[scale=0.27,angle=90]{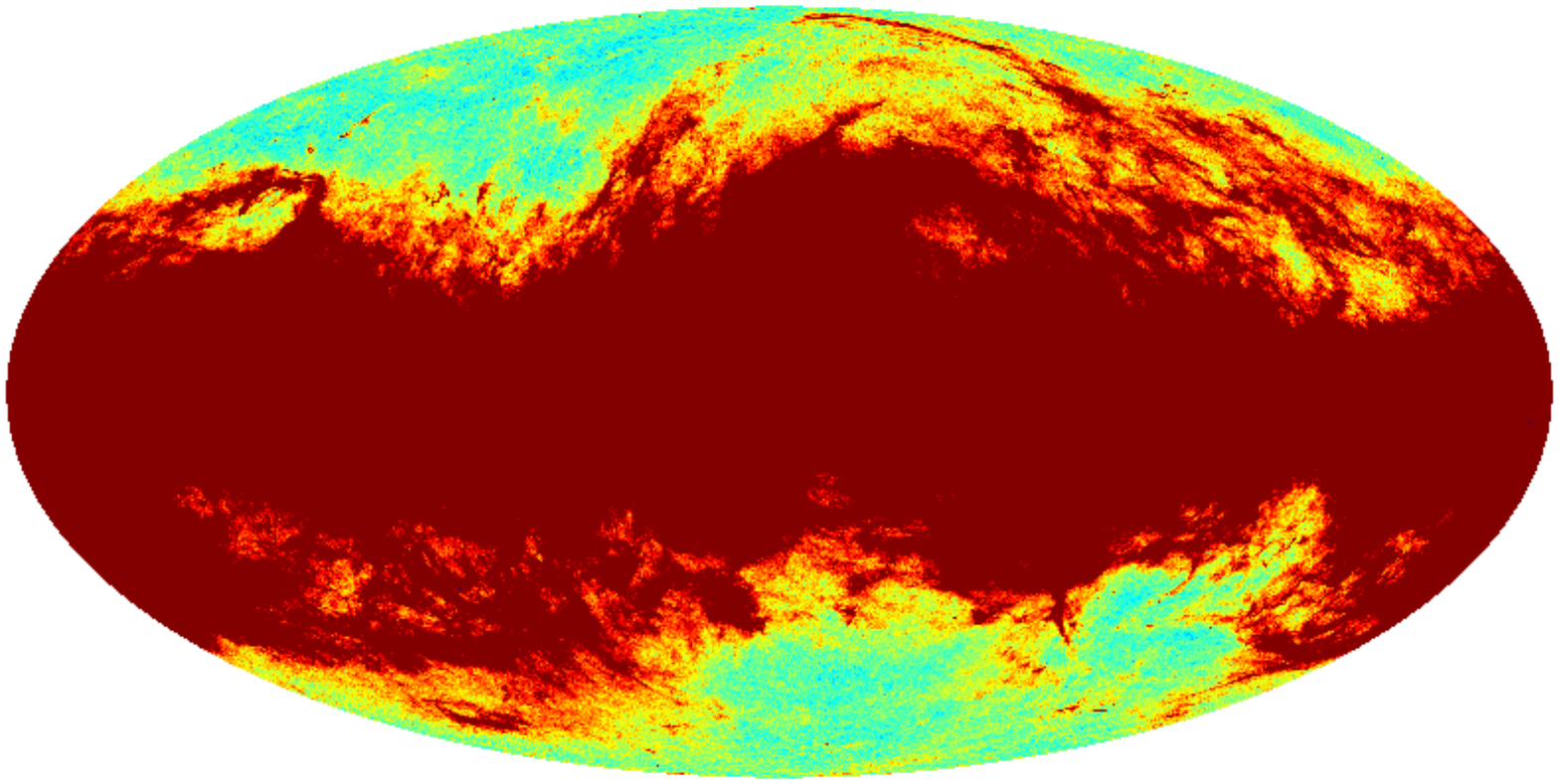}
\includegraphics[scale=0.27,angle=90]{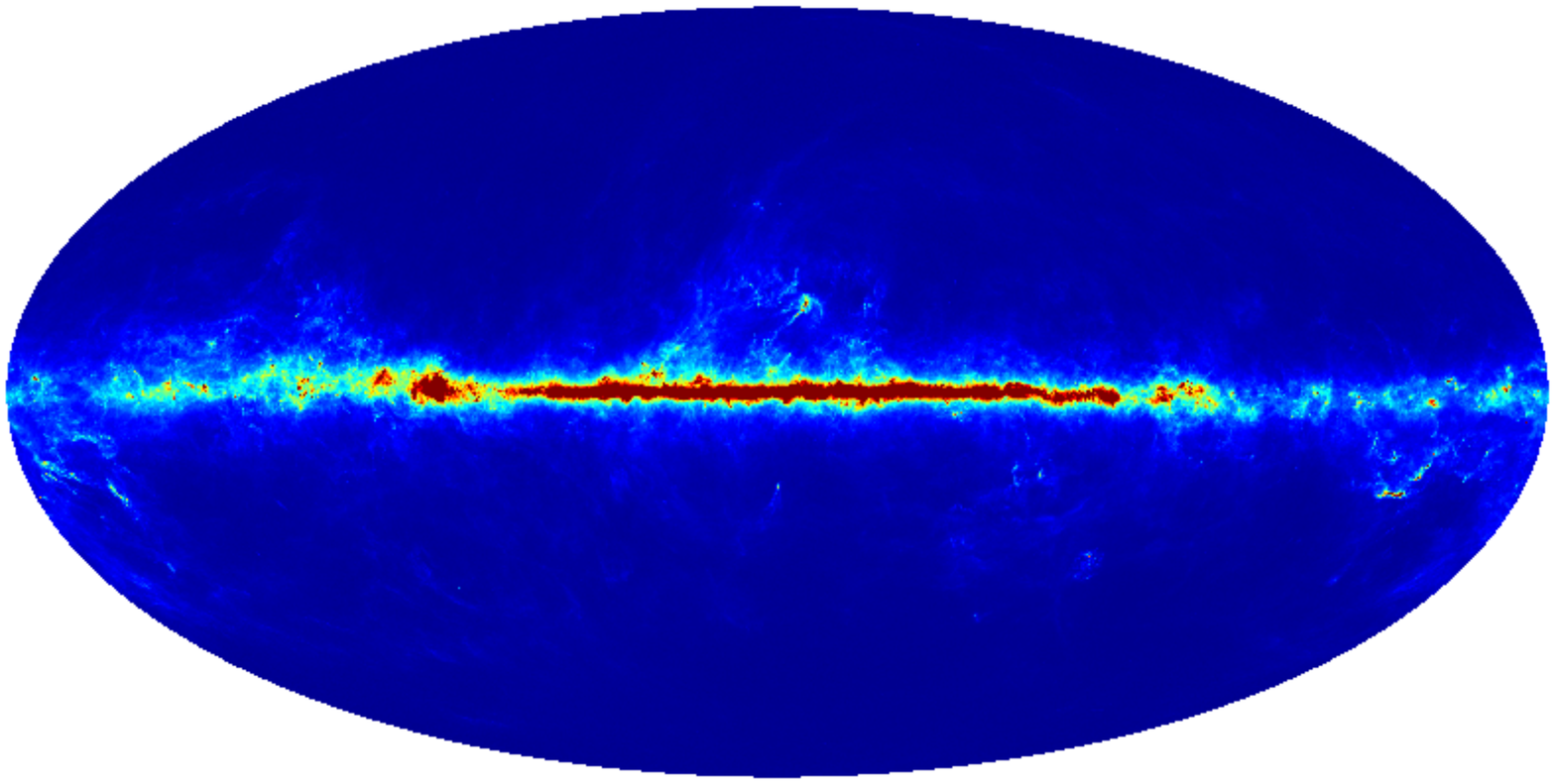}
\includegraphics[scale=0.27,angle=90]{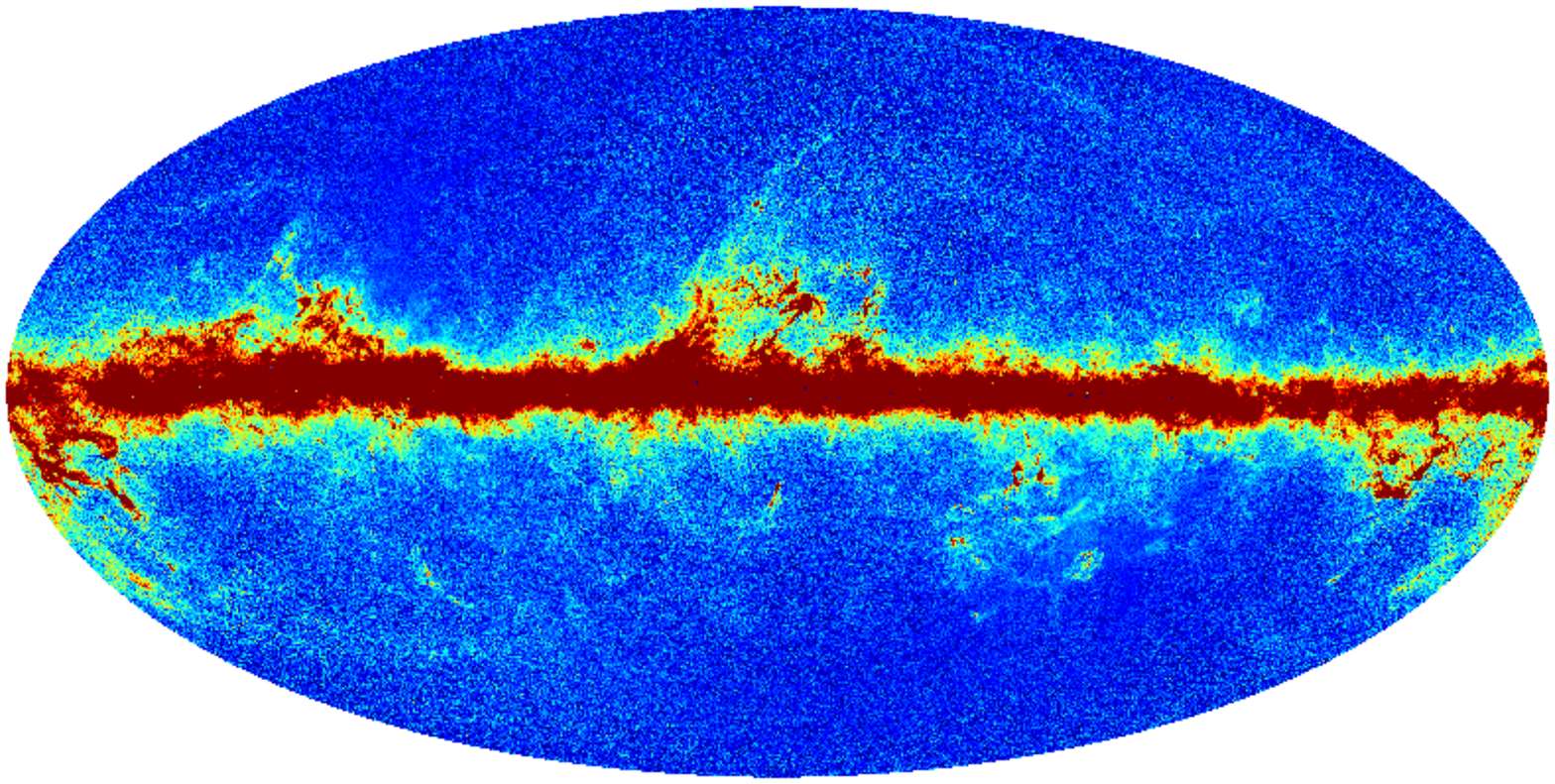}
\includegraphics[scale=0.27,angle=90]{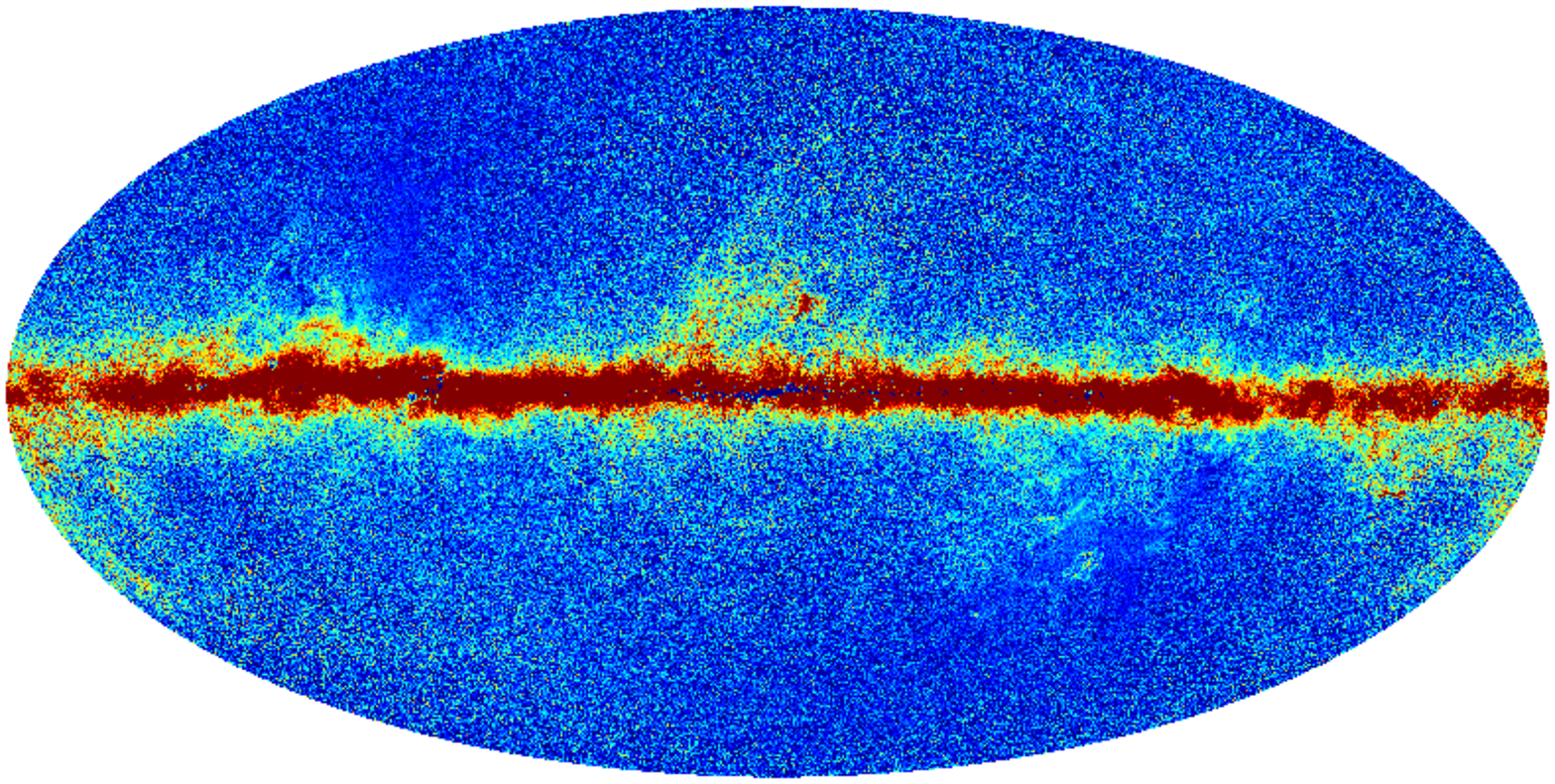}
\includegraphics[scale=0.27,angle=90]{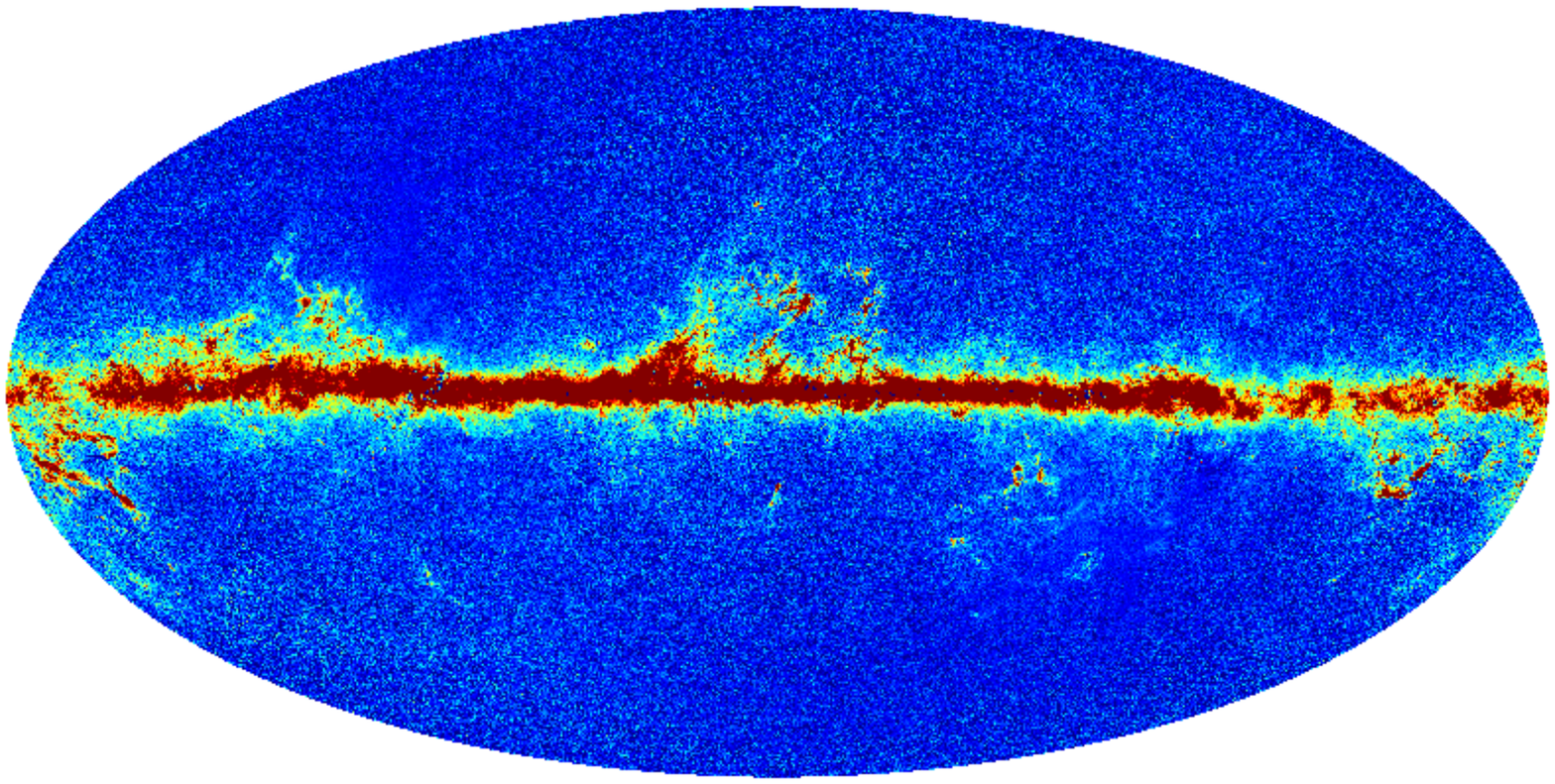}
\includegraphics[scale=0.27,angle=90]{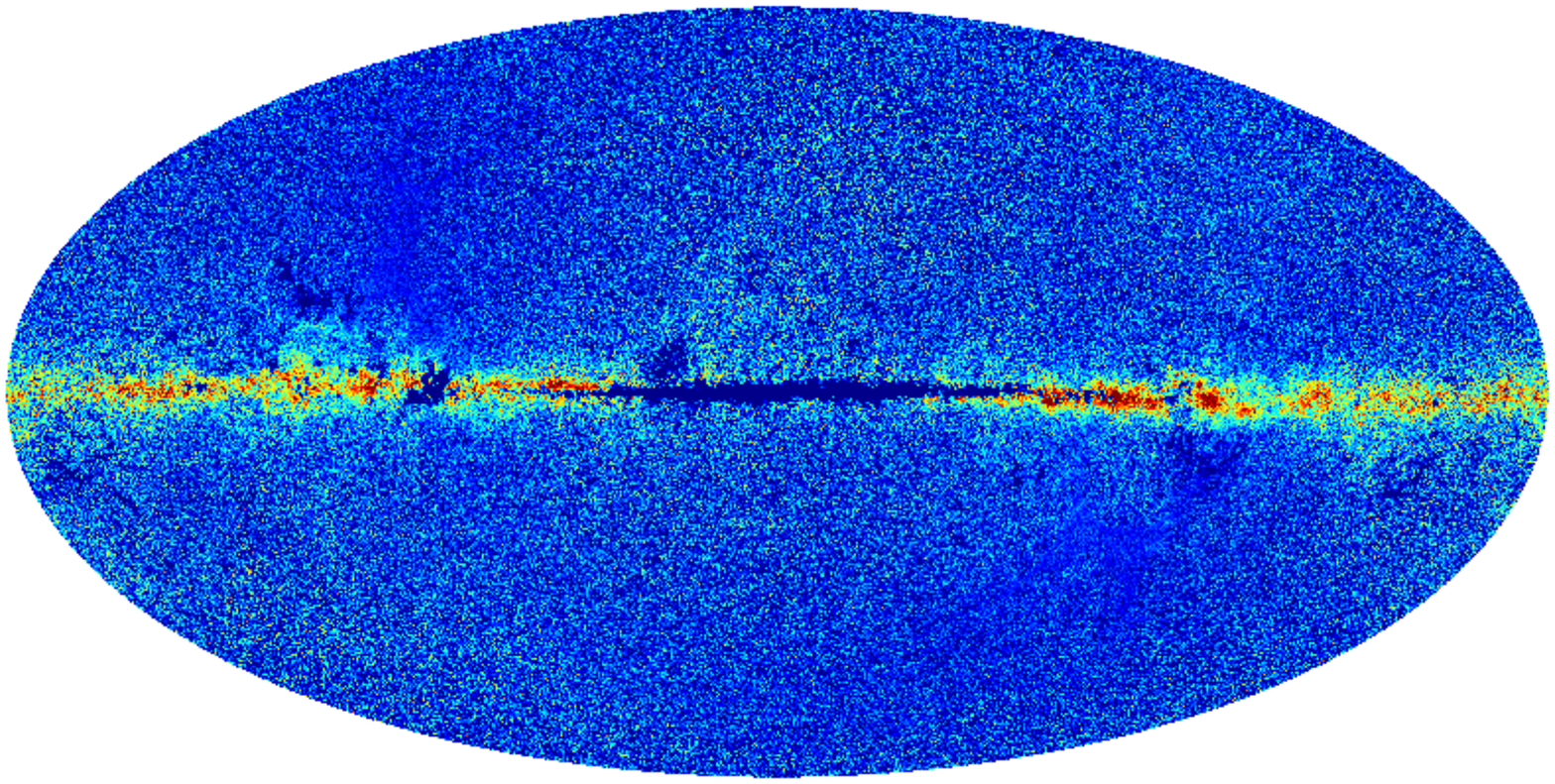}
\caption{\label{fig:maps}Maps of the Comptonization parameter, $y$, formed from linear combinations of the Planck HFI maps, shown on a scale $0 < y < 1 \times 10^{-4}$.  The residual contamination from foreground signals, primarly thermal dust emission, varies widely among the different maps, and would produce widely different results if it correlated with the lensing signal.  In all cases, the primary CMB fluctuations have been projected out by enforcing the condition $\sum_{\nu} b_{\nu} = 0$ (see text for details).  In addition, the maps shown here have: {\it top-left}  - the SZ signal retained, but no other spectra projected out (Planck A from Table \ref{tbl:var}).  {\it top-right} - both SZ signal and primary CMB projected out, but no other foreground spectra projected out (Planck E$'$).  {\it middle-left} - SZ signal retained, and dust projected out assuming $\beta_d = 1.8$ (Planck D).   {\it middle-right} - same as middle-left, but additionally projecting out CO emission (Planck D', see text for assumptions).   {\it lower-left} - Same as middle-left but with $\beta_d = 1.4$ (Planck B).   {\it lower-right} - same as middle-right, but with $\beta_d = 1.4$ (Planck B$'$).}
\end{center}
\end{figure*}

We form a given tSZ map, $y^i$, from a linear combination of band maps,
\begin{equation}
y^i(p) = \sum_\nu b^i_{\nu} \, T_{\nu}(p)/T_0,
\end{equation}
where the sum is over frequency band, $T_{\nu}(p)$ is the map at frequency $\nu$ and sky pixel $p$, and the $b^i_{\nu}$ are the coefficients of the $i$th estimate of a tSZ map, $y^i$.  The coefficients for the various trial maps are given in Table~\ref{tab:sz_coeff}, and several of the resulting maps are shown in Figure~\ref{fig:maps}.  In all cases, these coefficients satisfy the following constraints, 1) $\sum_{\nu} b_{\nu} S_{\rm SZ}(\nu) = 1$ to produce a map in units of the Comptonization parameter $y$; 2) $\sum_{\nu} b_{\nu} = 0$, to null the primary CMB fluctuations; 3)  $\sum_{\nu} b_{\nu} \cdot c_{\nu} (\nu/\nu_0)^{\beta_d} = 0$ to null dust emission with spectral index $\beta_d$ in antenna temperature units (the factor $c_{\nu}$ given in Table~\ref{tab:sz_coeff} converts antenna temperature to thermodynamic temperature), we consider a range of $\beta_d$ values in our tests.  Since we have four band maps available, we also have the option of imposing a 4th spectral constraint. We use this freedom to explore projecting out either radio emission or CO emission with an assumed set of line strength ratios.

The last column in Table~\ref{tab:sz_coeff} gives an indication of how much dust signal might remain in a given tSZ map.  The quoted value is the factor by which a dust signal with true index $\beta_d=2.0$ would be suppressed (or amplified) relative to its 100 GHz amplitude, in the given tSZ map.  For example, in map B, a dust signal with $\beta_d=2.0$ would survive with an amplitude of $-2$ times it 100 GHz amplitude.   If the cross-correlation we measure is dominated by dust emission with $\beta_d=2.0$ then the signal should scale like the numbers in this column.  Similar results are readily tabulated for other dust indices.  We place bounds on residual foreground contamination in the next section.

To carry out the cross-correlation, the convergence and Comptonization parameter maps, $\kappa(\thetavc_{ij})$ and $y(\thetavc_{ij})$ are pixelized on the same $\thetavc_{ij}$ grid. Note that the largest field, W1, is less than $10$ degrees on a side, therefore we can use the flat-sky approximation when performing Fourier analysis.

\section{Results}

Figure~\ref{SZkappa} shows the real space cross-correlation function $\xi_{y\kappa}(r)$ derived using Eq.~\ref{SZkappaXi}, for a variety of tSZ maps. The black circles show the cosmological signal and the red squares show the corresponding signal when the $B$-mode map is used for the mass map; the latter is consistent with zero. The errors were computed from $100$ mass map noise reconstructions, where the galaxies have been randomly rotated for each realization, leaving only the galaxy shape noise. The left two panels in Figure~\ref{SZkappa} show $\xi_{y\kappa}(r)$ using the $y$ maps A and I which contain the highest level of galactic and extra-galactic dust contamination; the four panels on the right show $\xi_{y\kappa}(r)$ using the cleaned $y$ maps listed in Table \ref{tbl:var}. Only cases C, E, G and H are shown because they correspond to the most extreme cleaning cases; the mass-tSZ cross-correlation using maps B, D and F are very similar. In all cases, $\xi_{y\kappa}(r)$ is relatively insensitive to the particular choice of $y$ map. 

The signal amplitude variable between the different cases are considerably less than the residual parameter $r_{2.0}$ given in Table \ref{tbl:var}. Even cases A and I, which show a larger amplitude, are consistent statistically with the other cases because the noise, caused by the uncleaned foreground and CIB emission, is much larger in the former. The significance of the detection is computed as follows: 

\begin{figure*}
\includegraphics[scale=0.85]{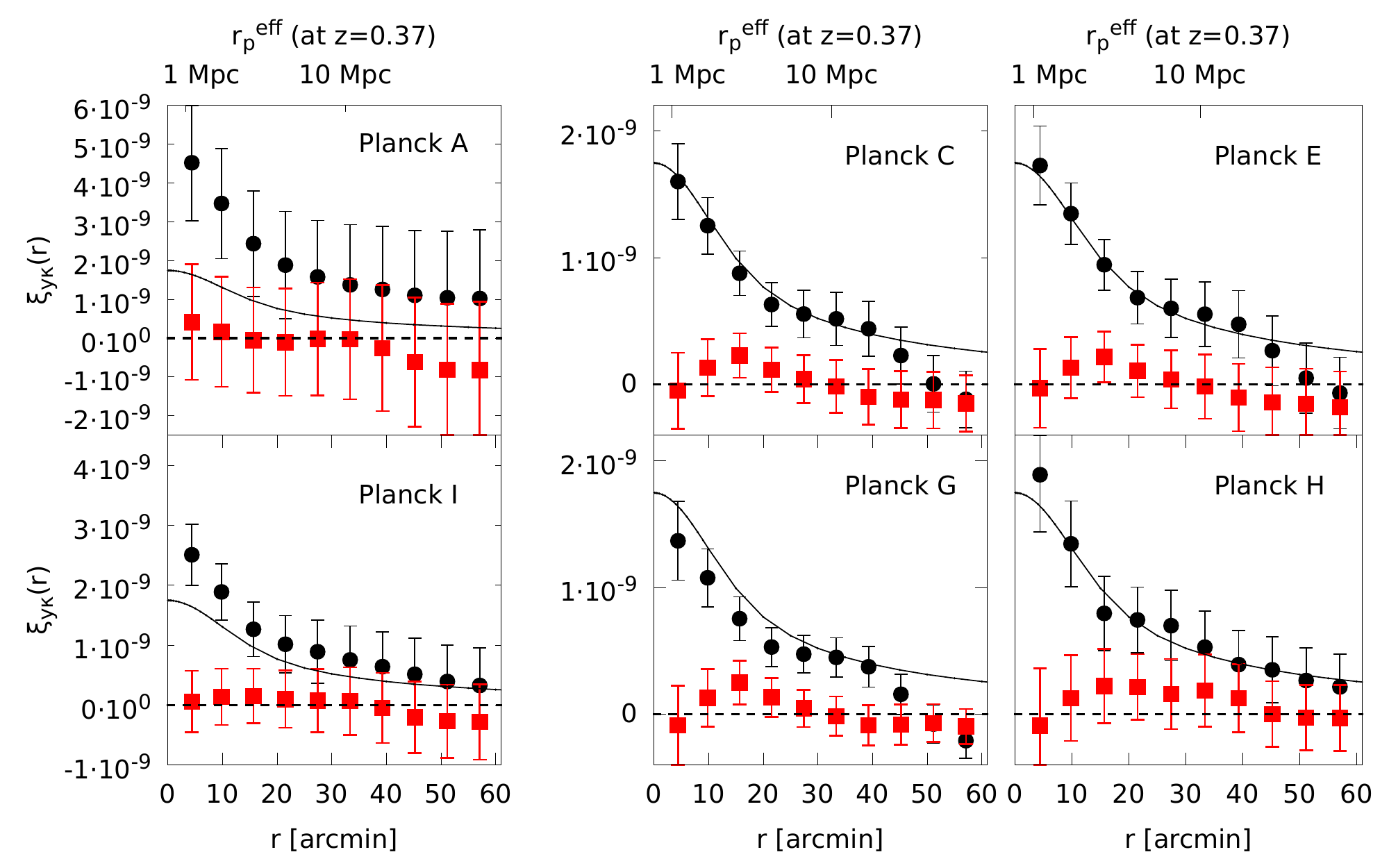}
\caption{\label{SZkappa} The filled black circles show the cross-correlation $\xi_{y\kappa}(r)$ between the gravitational lensing and the Sunyaev-Zeldovich maps. The bottom x-axis indicates the $y$-$\kappa$ pixels angular separation in arcminutes  and the top x-axis corresponds to the physical scale seen at the average redshift of the lenses. The filled red squares show the cross-correlation between SZ and the lensing $B$-mode. Planck A and I in the left two panels correspond to SZ maps with the largest foreground contamination (see Table \ref{tbl:var}). The four panels on the right (Planck C, E, G, H) correspond to the most extreme foreground cleaning parameters that we have considered (see Table \ref{tbl:var}). In all panels, the solid thick black line corresponds to the predicted $\xi_{y\kappa}(r)$ using WMAP7 cosmology and the source redshift distribution as shown in Figure \ref{nofz}, with an overall amplitude fit to the cases Planck B-H, as described in the text.}
\end{figure*}

The solid lines in Figure \ref{SZkappa} show the best-fit model derived from the average fit of the cross correlations with the Planck B-H maps. The model $\xi_{y\kappa}^{\rm model}(r)$ is given by Eq.~\ref{SZkappaXi} and $\xi_{y\kappa}^{\rm data}(r)$ is the data vector. The only free parameter in the model is the tSZ amplitude, $W^{\rm SZ}$, determined by Eq.\ref{WSZdef}; the cosmological parameters are set to the WMAP7 values with a cold dark matter power spectrum \citep{2011ApJS..192...18K}. The non-linear clustering is computed as per \citep{2003MNRAS.341.1311S}.  The likelihood function used for the fitting is given by:
\begin{equation}
\chi^2=-2\ln({\cal L})=\left(\xi_{y\kappa}^{\rm d}-\xi_{y\kappa}^{\rm m}\right)C^{-1}\left(\xi_{y\kappa}^{\rm d}-\xi_{y\kappa}^{\rm m}\right),
\end{equation}
where the covariance matrix $C$ is computed from the $100$ noise realizations and thus only includes a contribution from statistical noise.  For comparison, a noise covariance matrix was also derived from the data, but with only $4$ independent lensing fields, the variance estimate was not reliable, especially at the largest angular scale.  In particular, we found that the r.m.s among the four fields was roughly twice the statistical error on scales larger than $40$ arcmin, $\sim$50\% smaller for intermediate scales, and comparable for scales below $15$ arcmin.  We conclude that the error from the r.m.s. between the four fields is not reliable and use the statistical errors estimated from the $100$ noise reconstructions instead. We note that the cross-correlation derived from the $B$-mode map is statistically consistent with zero using this error estimate.  

The model amplitude, $W^{\rm SZ}$, is fit to the cross-correlations B-H listed in Table \ref{tbl:var}. The average amplitude is calculated from the $7$ cases and the quoted r.m.s. is the average of the individual r.m.s. values for each case.  We also quote a systematic error as the r.m.s. of the amplitude fitted for each case. The resulting constraint on the model amplitude is:
\begin{equation}
\left({b_{\rm gas}\over 1}\right) \left({T_e(0)\over 0.1 \textrm{ keV}}\right) \left({\bar{n}_e\over1 \textrm { m}^{-3}}\right) = 2.01\pm 0.31 \pm 0.21,
\label{SZresult}
\end{equation}
where the first error is statistical and the second is systematic.
Based on CMB and nucleosynthesis constraints, the total baryonic number density in the Universe is $\sim$5\% of the the critical density, or $n_{\rm b}\sim 0.5$ m$^{-3}$.  The gas bias on these scales is not well known, but numerical simulations suggest a value in the range of $4-9$ \citep{2000ApJ...531...31R,2001PhRvD..63f3001S}. Taking $b_{\rm gas}=6$ and $\bar{n}_e=0.25$ m$^{-3}$ we find $T_e \sim 0.13$ keV $\sim 10^6$ K.  Note that, with our $b_{\rm gas}=6$ assumption, if we take $\bar{n}_e$ six times larger, we find $T_e\sim 1.5\times 10^5$ K. Thus, if the signal we detect was all due to intercluster baryons in a warm plasma, it could account for {\em all} the missing baryons in the $10^5-10^7$ K temperature range. Confirming this will require more precise measurements and a proper halo model which includes a diffuse baryonic component.  While the three parameters, gas bias, temperature, and electron density are degenerate in our treatment, a more physically valid description of gas fluctuations would not allow one to increase $\bar{n}_e$ without also increasing $T_e(0)$ and $b_{\rm gas}$, which would inevitably change the constraints on these parameters. In future studies we will explore more realistic models beyond the constant bias case, which will help break the degeneracy between these parameters. 
\begin{figure}
\includegraphics[scale=0.43]{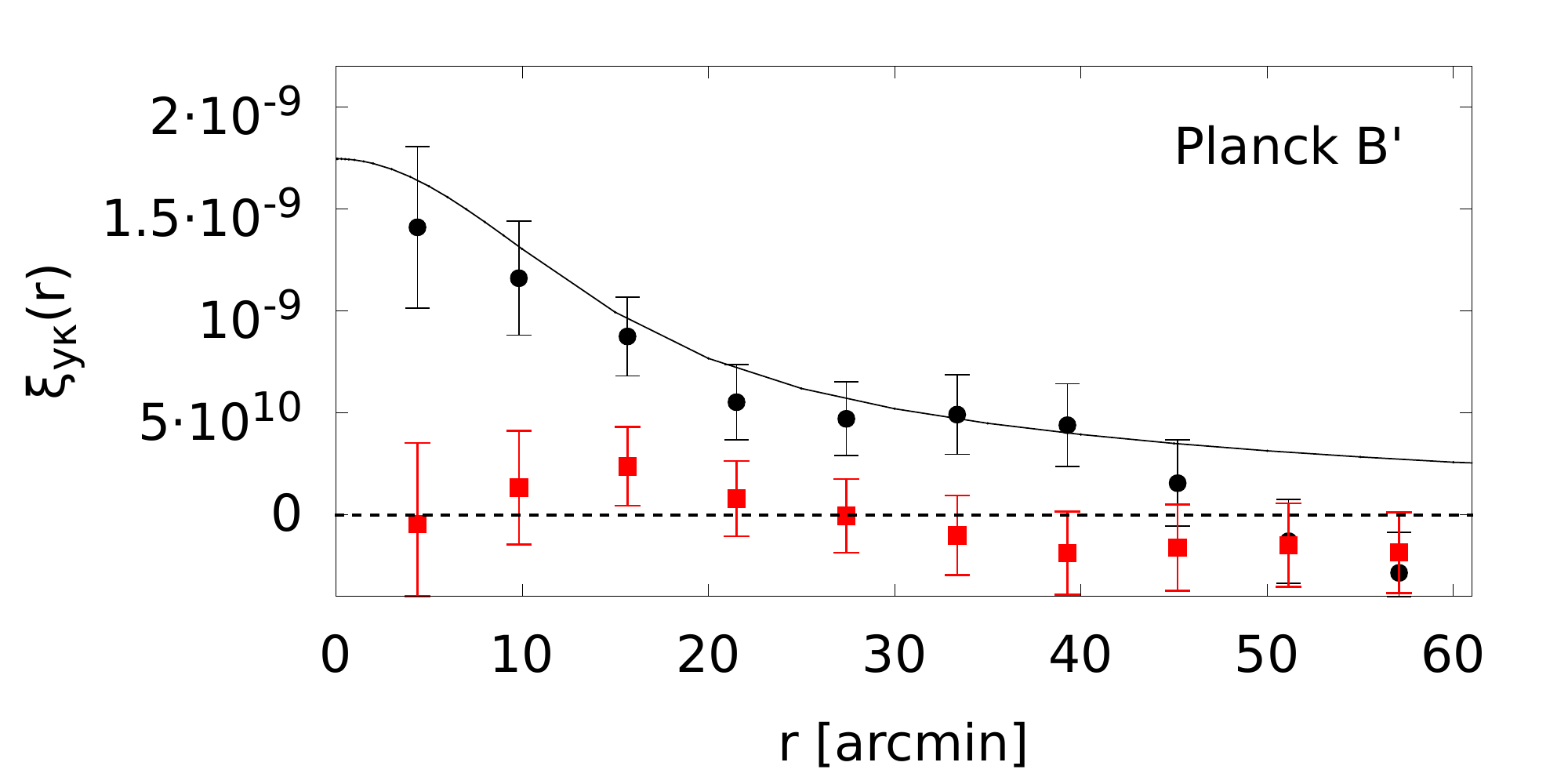}
\includegraphics[scale=0.38]{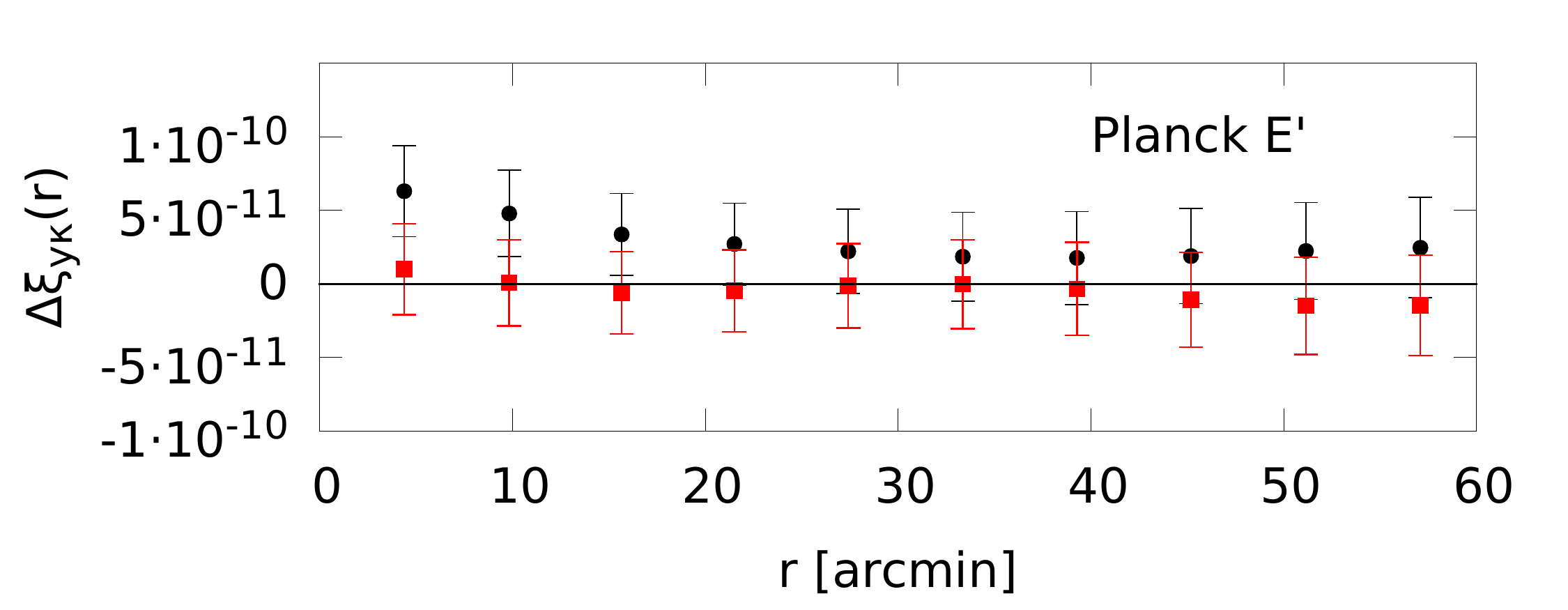}
\caption{\label{noymap}The top panel shows $\xi_{y\kappa}(r)$ when CO line emission is projected out in addition to a dust spectrum (see the text and Table~\ref{tab:sz_coeff} for details). The solid line is the same model as the one shown in Figure~\ref{SZkappa}. The bottom panel shows $\xi_{y\kappa}(r)$ when the tSZ signal is explicitly nulled in the $y$ map while a nominal 100 GHz dust signal is retained.  The dust contamination in map E$'$ should be comparable to the cases shown in the right panels of Figure~\ref{SZkappa}.}
\end{figure}
We note that gravitational lensing of the tSZ signal itself by foreground structures could potentially contaminate our signal. However, lensing itself cannot generate a correlation between tSZ and $\kappa$ if these two quantities are not physically correlated in the first place. This is because the gravitational lensing of a diffuse background does not add or suppress power, it only rearranges the power in $l$-bins. Using a line-of-sight approach similar to \citep{2006PhR...429....1L}, the contamination of the tSZ signal was predicted by \citep{2004MNRAS.349..816W} to be very small: $<3$\% at the arcminute scale, and much less at larger angular scales.

Figure \ref{noymap} shows two important additional null tests, which confirm that our signal is not significantly contaminated by residual foregrounds. The top panel shows $\xi_{y\kappa}(r)$ when CO emission is projected out, in addition to a dust component.  We assume a CO brightness ratio of 2:1 in the 217 and 100 GHz emission maps respectively (which encompass the J = 2-1 and J=1-0 transitions), in units of thermodynamic brightness temperature.  This is a typical ratio found in the outer galactic plane, which should be indicative of the range in our high-latitude lensing fields.  (The case shown in Figure~\ref{noymap} also assumes a dust spectrum $\beta_d=1.4$, but the cross-correlation results with CO projection are similarly insensitive to the projected dust index.)  The bottom panel of Figure~\ref{noymap} shows the cross-correlation with a map in which the tSZ signal is nulled but a dust signal -- scaled to a 100 GHz brightness temperature assuming $\beta_d=2.0$ -- is preserved (Planck E$'$ in Table \ref{tbl:var}, and the top-right panel of Figure~\ref{fig:maps}).  This map should retain the same level of dust contamination as most of the $y$ maps we studied above.

\section{Discussion}

We present the first direct measurement of a large-scale diffuse baryonic component in the universe using the cross-correlation between the thermal Sunyaev-Zeldovich (tSZ) map from the Planck satellite and the gravitational lensing mass map from the Canada France Hawaii Telescope Lensing Survey (CFHTLenS).  This is the first time a correlation between the tSZ signal and dark matter has been detected on cosmological scales.  Our approach differs from previous tSZ studies in that we do not stack the signal on known cluster positions, nor do we pre-select clusters in the lensing map \citep{2013arXiv1309.3282H}; the entire survey area contributes, including small groups, super-clusters, filaments and voids.

We perform a number of null tests to demonstrate that neither galactic and extragalactic dust emission, nor CO line emission, correlates with the lensing signal. The residual contamination is found to be less than $3\%$, which is a clear benefit of using a lensing signal that largely originates from redshifts $z<1$, where the extragalactic dust (CIB) signal is negligible \citep{2011A&A...536A..18P}.

A constant gas bias model, in which temperature fluctuations are neglected, provides a relatively good fit to the data. This result suggests that the diffuse component closely follows the dark matter distribution at least up to scales of $\sim$15 Mpc. With a gas bias of $6$, the constraint on the amplitude of the cross-correlation is consistent with a warm plasma of temperature $\sim$10$^6$ K, and an electron number density $\bar{n}_e\sim 0.25$ m$^{-3}$ (the latter corresponds to the missing baryon number density).  The cross-correlation extends to non-virialized regions, in agreement with predictions from hydrodynamical simulations \citep{2001ApJ...552..473D}.  A detailed analysis based on the halo model will be presented in a future study.  It will be particularly interesting to explore the possibility of constraining the physical mechanisms (e.g. outflows and accretion shocks) responsible for the ionization of these baryons.

Our approach is more direct, easier to implement, and less model dependent than indirect detection techniques relying on tracers of warm baryons \citep{2012ApJ...759...23S}. Its sensitivity to low gas density and temperature is also significantly better than for direct detection in the tSZ map \citep{2013A&A...550A.134P}. Existing and forthcoming lensing and tSZ surveys will improve measurement accuracy considerably.  High resolution tSZ data from ACT and SPT, when combined with available gravitational lensing data, will provide a unique probe of the warm plasma physics down to scales of clusters and groups of galaxies, while upcoming large-area gravitational lensing surveys (KiDS, DES), combined with Planck tSZ, will improve considerably the precision of our measurement at large scales. The combination of high-resolution and large scale correlation studies will ultimately give us a precise picture of the diffuse, ionized baryonic component in the Universe.

\section{acknowledgments}

\begin{acknowledgments}

LVW is supported by the  Natural Sciences and Engineering Research Council of Canada (NSERC) and the Canadian Institute for Advanced Research (CIfAR). GH and NM are supported by NSERC, CIfAR and by the Canada Research Chairs program. We thank Francois Bouchet for useful discussion about the Planck data and Yin-Zhe Ma for an independent check of some of the calculations performed in the paper.
We thank the CFHTLenS team for making their catalogues
publicly available to download from www.cfhtlens.org. This work was supported in part by the National Science Foundation under Grant No. PHYS-1066293 and the hospitality of the Aspen Center for Physics.

This work is partly based on observations obtained with Planck (http://www.esa.int/Planck), an ESA science mission with instruments and contributions directly funded by ESA Member States, NASA, and Canada.
This work is also partly based on observations obtained with
MegaPrime/MegaCam, a joint project of CFHT and CEA/IRFU,
at the Canada-France-Hawaii Telescope (CFHT) which is operated
by the National Research Council (NRC) of Canada, the Institut
National des Sciences de l’Univers of the Centre National
de la Recherche Scientifique (CNRS) of France, and the University
of Hawaii. This research used the facilities of the Canadian
Astronomy Data Centre operated by the National Research Council
of Canada with the support of the Canadian Space Agency.
CFHTLenS data processing was made possible thanks to significant
computing support from the NSERC Research Tools and Instruments
grant program.

\end{acknowledgments}

\bibliography{lensing}

\end{document}